# The Collaboration Gap in Human–AI Work

## Grounding and Repair Conditions for Stable Collaboration

Varad Vishwarupe · University of Oxford, United Kingdom · varad.vishwarupe@cs.ox.ac.uk
Marina Jirotka · University of Oxford, United Kingdom · marina.jirotkae@cs.ox.ac.uk
Nigel Shadbolt · University of Oxford, United Kingdom · nigel.shadbolt@cs.ox.ac.uk
Ivan Flechais · University of Oxford, United Kingdom · ivan.flechais@cs.ox.ac.uk

**Abstract.** LLMs are increasingly presented as collaborators in programming, design, writing, and analysis. Yet the practical experience of working with them often falls short of this promise. In many settings, users must diagnose misunderstandings, reconstruct missing assumptions, and repeatedly repair misaligned responses. This poster introduces a conceptual framework for understanding why such collaboration remains fragile. Drawing on a constructivist grounded-theory analysis of 16 interviews with designers, developers, and applied AI practitioners working on LLM-enabled systems, and informed by literature on human–AI collaboration, we argue that stable collaboration depends not only on model capability but on the interaction's grounding conditions. We distinguish three recurrent structures of human–AI work: one-shot assistance, weak collaboration with asymmetric repair, and grounded collaboration. We propose that collaboration breaks down when the appearance of partnership outpaces the grounding capacity of the interaction and contribute a framework for discussing grounding, repair, and interaction structure in LLM-enabled work.

## 1 Introduction

Large language models (LLMs) are increasingly integrated into professional workflows such as coding, writing, ideation, and technical analysis. In these settings, they are often described not only as tools, but as collaborators or teammates. Yet the practical experience of working with them frequently remains brittle. Users may obtain plausible outputs, but they also spend time re-prompting, correcting errors, reconstructing assumptions, and steering the interaction back on course.

This tension reflects a broader problem in collaborative AI. Davidson et al. show that models that perform strongly in isolation can degrade when they are required to collaborate, a phenomenon they term the *collaboration gap* (Davidson et al., 2025). In human–AI settings, adjacent problems have been observed in the difficulty users face when trying to interpret, guide, and productively rely on model outputs (Amershi et al., 2019; Bansal et al., 2021; Liao et al., 2020). The issue is not only whether outputs are correct. Collaboration fails when participants cannot reliably establish shared assumptions, interpret task state, or repair misunderstandings efficiently.

Varad Vishwarupe, Marina Jirotka, Nigel Shadbolt, Ivan Flechais. 2026. The Collaboration Gap in Human–AI Work: Grounding and Repair Conditions for Stable Collaboration. In: *Proceedings of the 24th EUSSET Conference on Computer-Supported Cooperative Work (ECSCW) – Posters and Demos*, Reports of the European Society for Socially Embedded Technologies. ISSN: 2510-2591 https://doi.org/10.48340/to-be-added

Find the latest version of this document in the EUSSET Digital Library: https://dl.eusset.eu/

This poster paper argues that such failures can be understood through the lens of *common ground*. In human communication, common ground refers to the shared beliefs, assumptions, goals, and situational awareness that enable coordinated action (Clark, 1996; Clark and Brennan, 1991; Roschelle and Teasley, 1995). Common ground is not simply present or absent. It is built through iterative acts of presentation, clarification, acknowledgment, and repair (Traum, 1998, 1992). Recent work suggests that these same mechanisms matter in human–AI collaboration, yet are often weakly supported in contemporary LLM interaction (Poelitz et al., 2026; Shaikh et al., 2024).

The paper asks a single guiding question:

- **How can we explain the fragility of human–AI collaboration in terms of grounding and repair?**

We answer this by proposing a compact framework that distinguishes three interaction structures—one-shot assistance, weak collaboration, and grounded collaboration—and by showing how these differ in grounding capacity and repair burden.

## 2  Motivation and Background

Research in HCI has long emphasized that effective human–AI interaction depends on more than raw system performance. Appropriate reliance, transparency, and usable explanation all shape whether people can work productively with AI systems (Amershi et al., 2019; Eiband et al., 2018; Hoff and Bashir, 2015; Kaur et al., 2020; ?). At the same time, recent work on human–AI collaboration suggests that coordination itself must be treated as a distinct challenge. Davidson et al. show that models which perform well alone may perform poorly when required to align representations, exchange partial information, and co-develop solutions (Davidson et al., 2025). Poelitz et al. similarly argue that effective human–AI collaboration depends on establishing and maintaining common ground, and show that current systems often exhibit overly long one-shot responses, shallow grounding behaviour, and asymmetric effort that leaves the burden of alignment on the human (Poelitz et al., 2026).

The concept of common ground provides a useful foundation here. Clark and Brennan define grounding as the process by which participants establish sufficient mutual understanding for current purposes (Clark and Brennan, 1991). Classic studies in communication and CSCW show that grounding depends on feedback, shared references, repair, and access to evolving task state (Clark and Wilkes-Gibbs, 1986; Fussell and Krauss, 1992; Gergle et al., 2013; Kraut et al., 2002). These processes become especially important when collaboration requires joint interpretation rather than simple task execution.

In parallel, recent HCI work has shown that practitioners already reason about LLM use in sociotechnical terms. Rather than asking only whether a model is accurate, they ask what role it should occupy relative to human work, what forms of oversight remain possible, and how responsibility for outcomes is distributed (Shneiderman, 2020; Vishwarupe et al., 2026). This makes grounding and repair design questions, not merely model questions.

# 3 Study Basis

The framework presented in this poster is informed by a constructivist grounded-theory analysis of 16 semi-structured interviews with designers, developers, and applied AI practitioners working on LLM-enabled systems in large technology organisations. Participants reflected on how LLMs were used in workflows involving drafting, ideation, coding, evaluation, and decision support. They described when collaboration with the model felt productive, when it became fragile, and how they responded when outputs diverged from task requirements or expectations.

The analysis was abductive. Categories emerged through iterative coding of practitioner accounts, while being sensitised by prior work on common ground, repair, and human–AI collaboration (Charmaz, 2014; Clark and Brennan, 1991; Poelitz et al., 2026). The goal was not to evaluate model performance directly, but to understand how practitioners recognised and responded to breakdowns in collaborative interaction.

# 4 Framework: Grounding and Repair Conditions

Figure 1 presents the core framework. It proposes that stable human–AI collaboration depends on the relationship between *grounding capacity* and *repair burden distribution*. Grounding capacity refers to the extent to which an interaction supports shared understanding of assumptions, context, task state, and next actions. Repair burden distribution refers to who must do the work of detecting and correcting misalignment.

The framework distinguishes three recurrent interaction structures.

## 4.1 One-shot assistance

At the lowest level of grounding, interaction remains structurally close to request-and-response assistance. The user provides a prompt, the system produces an output, and any correction happens post hoc. Shared understanding remains low and largely implicit. This structure can work for bounded, low-risk tasks such as summarisation or boilerplate generation, but it does not support deeper collaboration because the system does not help maintain a shared problem representation.

## 4.2 Weak collaboration

A second structure emerges when interaction becomes iterative. Users refine prompts, correct outputs, add context, or ask for revisions. These interactions may appear collaborative, but the burden of repair remains largely human. The user must infer what has gone wrong, reconstruct missing assumptions, and guide the system back toward the task. In this sense, weak collaboration sits at the centre of the collaboration gap: the interaction looks like partnership, but grounding remains partial and repair remains asymmetric.

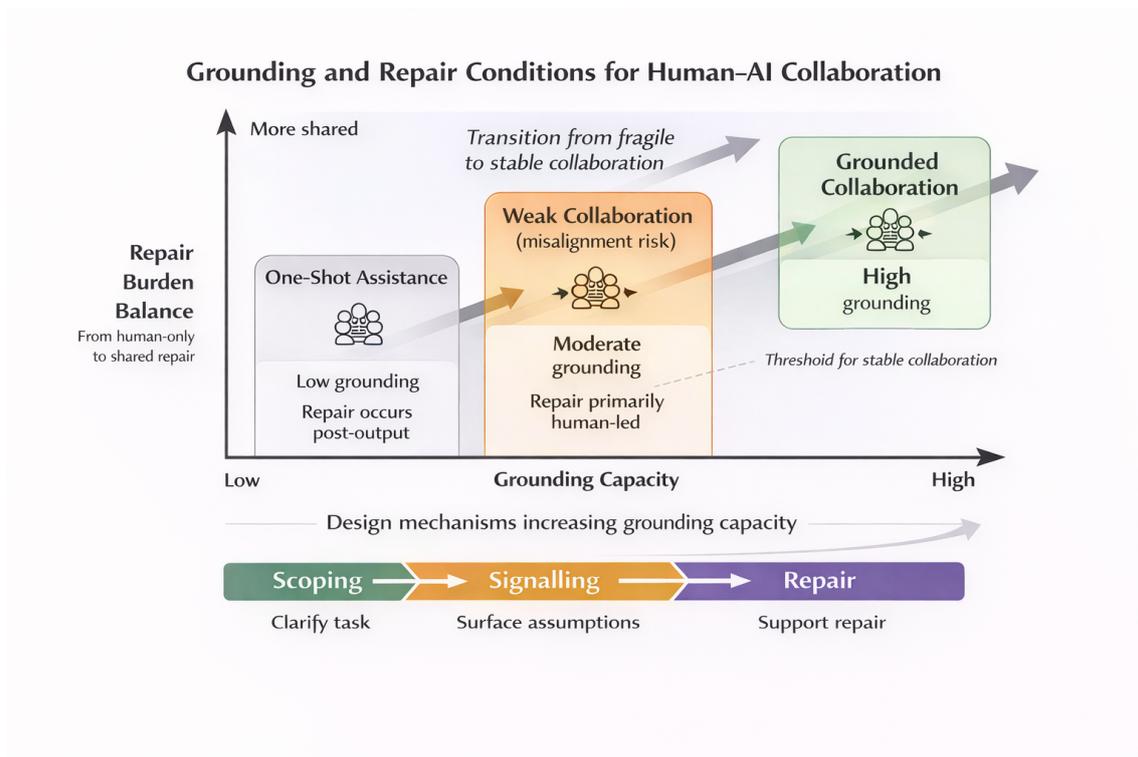

**Figure 1.** Grounding and repair conditions for human–AI collaboration. One-shot assistance, weak collaboration, and grounded collaboration differ in how much grounding the interaction supports and how repair burden is distributed.

## 4.3 Grounded collaboration

At the highest level of grounding, the interaction begins to support explicit clarification, signalling, and mutual repair. The system helps surface assumptions, track context, and make misalignment more visible. Final authority may still remain with the human, but the interaction itself becomes more balanced because repair no longer depends entirely on human improvisation. This is the condition under which collaboration becomes more stable.

# 5 Design Mechanisms

The practitioner interviews further suggest three recurring mechanisms through which teams try to increase grounding capacity.

**SCOPING.** Teams reduce ambiguity by narrowing task scope, fixing formats, and breaking work into smaller stages. Scoping limits what the interaction must jointly maintain.

**SIGNALLING.** Teams seek visible evidence of what the system has understood by asking it to restate assumptions, summarise task state, or surface uncertainty. Signalling makes shared understanding more inspectable.

**REPAIR.** Teams create explicit pathways for revision, rollback, clarification, and contestation. Repair becomes a design feature rather than an afterthought.

These mechanisms are included in Figure 1 because they are not merely interface embellishments. They are conditions that increase the interaction's capacity to sustain collaboration.

# 6 Implications

This framework suggests three implications for CSCW and HCI research on LLM-enabled work.

First, collaboration should not be inferred from interactivity alone. Multi-turn interaction does not necessarily imply grounded collaboration. Weak collaboration may still leave the burden of alignment almost entirely with the user.

Second, the collaboration gap is not only a model capability problem. It is also a workflow and interaction design problem. Seemingly useful systems often fail because they do not provide enough support for shared understanding, visibility, and repair.

Third, collaboration with LLMs should be evaluated in terms of *who carries repair burden*. This provides a practical indicator of where collaboration actually sits: post-hoc human correction, human-heavy iterative steering, or more mutual and inspectable interaction.

**Table 1.** Properties of the three interaction structures observed in practitioner accounts of LLM-enabled work.

| Interaction structure | Typical task setting | Shared understanding | Repair burden |
|---|---|---|---|
| One-shot assistance | Summarisation, rewriting, boilerplate generation | Low and largely implicit | Human, after output |
| Weak collaboration | Iterative drafting, debugging, exploratory prompting | Moderate but unstable or skewed | Primarily human |
| Grounded collaboration | Staged reasoning, bounded exploration, structured review | More explicit and inspectable | Shared and reduced |

# 7 Conclusion

Drawing on practitioner accounts and grounding theory, this paper argues that stable collaboration depends on grounding capacity and repair burden distribution. The framework distinguishes three structures–one-shot assistance, weak collaboration, and grounded collaboration–and highlights scoping, signalling, and repair as key mechanisms for workable interaction. By reframing the collaboration gap as a grounding and repair problem, this work offers a conceptual lens for rethinking how human–AI collaboration is designed.